\documentclass[%
 reprint,
 amsmath,amssymb,
 aps,
 pra,
superscriptaddress]{revtex4-2}

\UseRawInputEncoding

\usepackage[colorlinks=true, allcolors=blue]{hyperref}
\usepackage{amsmath}
\usepackage{braket}
\usepackage{graphicx}
\usepackage{dcolumn}
\usepackage{bm}
\usepackage{listings}
\usepackage{xcolor}

\begin{document}
\preprint{APS/123-QED}
\title{HOBOTAN: Efficient Higher Order Binary Optimization
Solver with Tensor Networks and PyTorch}
\author{Shoya Yasuda}
\email{yasuda@vigne-cla.com}
\affiliation{Vignette \& Clarity, Inc., 6-24-2 Honkomagome, Bunkyo-ku, Tokyo, 113-0021, Japan}
\affiliation{Tokyo Institute of Technology, School of Computing, J2 bldg. room1710, 4259 Nagatsuta-cho, Midori-ku, Yokohama, 226-8503, Japan}
\author{Shunsuke Sotobayashi}
\email{derwind0707@gmail.com}
\affiliation{Independent Researcher}
\author{Yuichiro Minato}
\email{minato@blueqat.com}
\affiliation{blueqat Inc., 2-24-12 Shibuya, Shibuya-ku, Tokyo 150-6139, Japan}
\begin{abstract}
In this study, we introduce HOBOTAN, a new solver designed for Higher Order Binary Optimization (HOBO). HOBOTAN supports both CPU and GPU, with the GPU version developed based on PyTorch, offering a fast and scalable system. This solver utilizes tensor networks to solve combinatorial optimization problems, employing a HOBO tensor that maps the problem and performs tensor contractions as needed. Additionally, by combining techniques such as batch processing for tensor optimization and binary-based integer encoding, we significantly enhance the efficiency of combinatorial optimization. In the future, the utilization of increased GPU numbers is expected to harness greater computational power, enabling efficient collaboration between multiple GPUs for high scalability. Moreover, HOBOTAN is designed within the framework of quantum computing, thus providing insights for future quantum computer applications. This paper details the design, implementation, performance evaluation, and scalability of HOBOTAN, demonstrating its effectiveness.
\end{abstract}
\maketitle
\section{Introduction}
Combinatorial optimization problems play a crucial role in various fields such as logistics~\cite{Domino_2022}, finance, healthcare, and energy management. Solving these problems requires advanced computational resources~\cite{Suksmono2022} and efficient algorithms.

In this study, we developed a new solver called HOBOTAN, which supports Higher Order Binary Optimization (HOBO). HOBOTAN is designed to solve high-order combinatorial optimization problems using tensor networks~\cite{minato2024tensornetworkbasedhobo} and supports both CPU and GPU. The GPU version, based on PyTorch, is particularly designed to be fast and scalable, showing high performance even for large optimization problems.

The core feature of HOBOTAN is the preparation of a HOBO tensor that maps the problem and performs tensor contractions as needed. This enables efficient solving of complex combinatorial optimization problems. Furthermore, by combining batch processing for tensor optimization and binary-based integer encoding techniques, we significantly extend the advantages of combinatorial optimization.

In the future, increasing the number of GPUs is expected to harness greater computational power, allowing for high scalability by efficiently coordinating multiple GPUs. This will enable HOBOTAN to handle even larger optimization problems beyond the current capabilities of high-performance computing.

Additionally, HOBOTAN is designed within the framework of quantum computing, providing insights for future quantum computer applications. When quantum computers become practical, the optimization techniques using tensor networks developed in HOBOTAN can potentially be applied directly. The knowledge gained from the development of HOBOTAN is expected to contribute to research in the field of quantum computing.

This paper first details the design and implementation of HOBOTAN, followed by performance evaluation through some benchmark tests. Finally, we discuss the new possibilities HOBOTAN brings to combinatorial optimization and outline future research challenges and prospects.

\section{Theory}

\subsection{Overview of Tensor Networks}

Tensor networks~\cite{wang2023tensornetworksmeetneural} are a mathematical framework designed to efficiently handle complex operations on high-dimensional tensors. A tensor is a multi-dimensional array, generalizing scalars (0-dimensional), vectors (1-dimensional), and matrices (2-dimensional). Tensor networks reduce computational complexity by contracting tensor elements. Contraction is an operation that sums over specific dimensions of tensors, thus aggregating into smaller tensors while maintaining the overall relationships within the network.

\subsection{Formulation of QUBO and HOBO in Combinatorial Optimization Problems}

Combinatorial optimization problems involve finding a combination of variables that minimizes (or maximizes) a given cost function~\cite{Andrew2014}. Quadratic Unconstrained Binary Optimization (QUBO) is formulated as follows:

\[
f(x) = \sum_{i} a_i x_i + \sum_{i < j} b_{ij} x_i x_j
\]

where \( x_i \) are binary variables (0 or 1), and \( a_i \) and \( b_{ij} \) are coefficients.

Higher Order Binary Optimization (HOBO) generalizes QUBO by including higher-order terms. HOBO is formulated as follows:

\[
f(x) = \sum_{i} a_i x_i + \sum_{i < j} b_{ij} x_i x_j + \sum_{i < j < k} c_{ijk} x_i x_j x_k + \cdots
\]

where \( c_{ijk} \) are third-order coefficients, and higher-order terms are included.

\subsection{Formulating HOBO as Multi-dimensional Tensors and Implementing Contractions with Vector x}

Based on the HOBO formulation, complex combinatorial optimization problems with higher-order terms can be implemented using tensor networks. First, prepare the tensor (HOBO tensor) representing the HOBO problem, where each element of the tensor corresponds to a coefficient of the problem.

Next, define the binary variables \( x \) as a vector and perform contractions with the HOBO tensor. Contraction sums over the dimensions of the tensor and the elements of the vector, calculating the cost function value.

Specifically, this can be formulated as:

\[
H(x) = \sum_{i,j,k,\ldots} H_{ijk\ldots} x_i x_j x_k \ldots
\]

where \( H_{ijk\ldots} \) are the elements of the HOBO tensor, and \( x_i, x_j, x_k, \ldots \) are binary variables. By contracting tensor \( H \) with vector \( x \), the cost function value is computed.

The implementation steps are as follows:
\begin{enumerate}
\item Prepare the tensor \( H \) representing the problem.
\item Define the binary variable vector \( x \).
\item Contract tensor \( H \) with vector \( x \) to compute the cost function value.
\end{enumerate}

\subsection{Methods for Minimizing or Maximizing the Cost Function}

After computing the cost function value, various optimization algorithms can be used to minimize or maximize the cost function, thus solving the combinatorial optimization problem. The following are some representative methods:

\subsubsection{Simulated Annealing}

Simulated annealing is a probabilistic optimization algorithm inspired by the annealing process in physics. It starts at a high temperature and gradually cools down, reducing the search space. In the initial stages, a wide search prevents getting stuck in local optima, eventually approaching the global optimum. Simulated annealing can efficiently find optimal solutions from a broad solution space for HOBO problems.

\subsubsection{Gradient Descent}

Gradient descent is an optimization method that uses the gradient (slope) of the cost function to minimize its value. The gradient of the cost function is computed, and variables are updated along the gradient direction to converge to the optimum. For HOBO problems, the gradient is computed based on tensor contraction results, and variables are updated iteratively. Variants like mini-batch gradient descent can improve computational efficiency for large-scale problems.

These methods enable efficient optimization of HOBO problems using tensor networks. Thus, HOBOTAN serves as a powerful tool for solving complex combinatorial optimization problems.

\section{HOBOTAN Implementation}

\subsection{Overview of Einsum Calculation}

Einsum (Einstein summation convention) is a powerful notation used for tensor operations, particularly in tensor network calculations. It allows for concise and efficient representation of complex tensor contractions, which are crucial for solving high-dimensional optimization problems. Einsum simplifies the expression of these operations by summing over repeated indices in a product of tensors.

For example, the matrix multiplication \( C = AB \) can be expressed using Einsum as follows:

\[
C_{ij} = \sum_{k} A_{ik} B_{kj}
\]

This efficiently performs tensor contractions, simplifying the description of tensor calculations and improving computational efficiency.

\subsection{Utilizing PyTorch for Combinatorial Optimization}

Considering future scalability, especially leveraging GPU capabilities, we utilize PyTorch, a machine learning framework, for combinatorial optimization calculations. By using PyTorch as the backend, we benefit from parallel computation and automatic differentiation, which are advantageous for solving optimization problems. In PyTorch, tensors are created using \texttt{torch.tensor}, and Einsum calculations are performed using \texttt{torch.einsum}.

\subsection{Conversion to HOBO Tensor}

To solve HOBO problems, the problem must be converted into a HOBO tensor. Specifically, combinatorial optimization problems with high-order terms are transformed into tensors of appropriate dimensions. In this conversion, it is crucial to replicate the smallest subscript number to extend the dimensions.

For example, converting a 3-dimensional tensor term \( c_{ijk} x_i x_j x_k \) into a 5-dimensional HOBO tensor involves replicating the smallest subscript number to extend the dimensions as follows:

\[
c_{ijk} x_i x_j x_k \rightarrow c_{iiijk} x_i x_i x_i x_j x_k
\]

\section{Special Implementations in HOBOTAN}

\subsection{Compilation of Formulation}

As mentioned in the previous chapter, if the order of terms does not reach the maximum order of the HOBO tensor, the compiler automatically replicates the smallest order to store the coefficients in the HOBO tensor. Specifically, when converting a 3-dimensional term \(c_{ijk}x_ix_jx_k\) into a 5-dimensional HOBO tensor, the smallest subscript number is replicated to extend the dimensions. This results in the following transformation:

\[
c_{ijk} x_i x_j x_k \rightarrow c_{iiijk} x_i x_i x_i x_j x_k
\]

This transformation ensures that the HOBO tensor has consistent dimensions and can appropriately represent all terms.

\subsection{Integer Encoding}

Using HOBO allows for integer encoding from binary encoding, significantly reducing the number of qubits required for the algorithm. This method defines a new integer variable \(y_i\) using the original binary variables \(x_i\). Specifically, this is expressed as:

\[
y_i = \sum_{k=0}^{n} 2^k x_{ik}
\]

where \(x_{ik}\) are binary variables and \(y_i\) takes integer values. This encoding enhances the expressive power of the problem and reduces the required number of qubits.

\subsection{Tensor Batch System}

For cost calculation in HOBO (including QUBO), we use tensor contraction through einsum. Typically, tensor contraction calculates the cost as a scalar by contracting with the corresponding vector \( x \). In this implementation, to fully utilize GPU performance and achieve parallelization, we prepare multiple vectors \( x \), expand them into a matrix, and contract them directly with the HOBO tensor. For example, if the matrix \( X \) contains multiple vectors \( x \), the cost can be calculated as follows:

\[
E = \sum_{i,j,k,\ldots} H_{ijk\ldots} X_{i} X_{j} X_{k} \ldots
\]

This method allows us to simultaneously compute the cost corresponding to multiple vectors. This approach supports multiple sampling typically executed by heuristic algorithms, ensuring parallelism and speeding up computation through tensor network-based techniques.

\subsection{Tensor Dimension Reduction and Efficiency Through Tensor Networks}

The core HOBO tensor of this algorithm is computationally intensive. By optimizing this configuration, we can significantly reduce the computational load. Specifically, unnecessary dimensions are reduced, and tensor network optimization is performed to improve computational efficiency.

\section{Solving Specific HOBO Problems}

Next, we demonstrate how to solve specific HOBO problems using the solver. Three examples are provided here.

\subsection{Problem 1: Maximizing the Number of Students Seated in a 5 by 5 Grid}

In this problem, students cannot be seated in three consecutive seats either in a row or in a column, which is a 3rd-order constraint.

\subsubsection{Implementation}

\begin{lstlisting}[language=Python]
import numpy as np
from hobotan import *
import matplotlib.pyplot as plt

# Define quantum bits
q = symbols_list([5, 5], 'q{}_{}')

# Objective: Preferably occupy all seats
H1 = 0
for i in range(5):
    for j in range(5):
        H1 += - q[i, j]

# Constraint: No three consecutive seats in any row or column
H2 = 0
for i in range(5):
    for j in range(5 - 3 + 1):
        H2 += np.prod(q[i, j:j+3])
for j in range(5):
    for i in range(5 - 3 + 1):
        H2 += np.prod(q[i:i+3, j])

# Combine terms
H = H1 + 10 * H2

# Compile into HOBO tensor
hobo, offset = Compile(H).get_hobo()
print(f'offset\n{offset}')

# Select sampler
solver = sampler.MIKASAmpler()

# Perform sampling
result = solver.run(hobo, shots=10000)

# Display top 3 results
for r in result[:3]:
    print(f'Energy {r[1]}, Occurrence {r[2]}')

    # Convert result to array
    arr, subs = Auto_array(r[0]).get_ndarray('q{}_{}')
    print(arr)

    # Visualize result
    img, subs = Auto_array(r[0]).get_image('q{}_{}')
    plt.figure(figsize=(2, 2))
    plt.imshow(img)
    plt.show()
\end{lstlisting}

\subsubsection{Results}

\begin{verbatim}
offset
0
MODE: GPU
DEVICE: cuda:0
Energy -17.0, Occurrence 844
[[1 1 0 1 1]
 [1 1 0 1 1]
 [0 0 1 0 0]
 [1 1 0 1 1]
 [1 1 0 1 1]]
Energy -17.0, Occurrence 498
[[1 1 0 1 1]
 [0 1 1 0 1]
 [1 0 1 1 0]
 [1 1 0 1 1]
 [0 1 1 0 1]]
Energy -17.0, Occurrence 538
[[1 1 0 1 1]
 [1 0 1 1 0]
 [0 1 1 0 1]
 [1 1 0 1 1]
 [1 0 1 1 0]]
\end{verbatim}

The solver successfully identified configurations that maximize the number of students seated while adhering to the constraint of not having three consecutive seats occupied in any row or column. By introducing the tensor batch system, the solver efficiently handles real-world problems, allowing it to solve even 10,000 samples rapidly. This demonstrates the effectiveness and utility of the HOBO solver in tackling combinatorial optimization problems with higher-order constraints.

\begin{figure}[h!]
\centering
\includegraphics[width=0.3\textwidth]{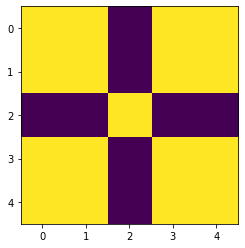}
\caption{Configuration 1: Energy -17.0, Occurrence 844}
\end{figure}

\begin{figure}[h!]
\centering
\includegraphics[width=0.3\textwidth]{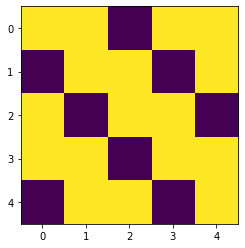}
\caption{Configuration 2: Energy -17.0, Occurrence 498}
\end{figure}

\begin{figure}[h!]
\centering
\includegraphics[width=0.3\textwidth]{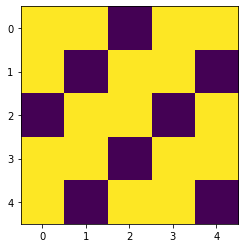}
\caption{Configuration 3: Energy -17.0, Occurrence 538}
\end{figure}

\subsection{Problem 2: Finding Pythagorean Triples}

Next, we solve the problem of finding all sets of three integers that satisfy the relationship \(a^2 + b^2 = c^2\), known as Pythagorean triples.

\subsubsection{Implementation}

\begin{lstlisting}[language=Python]
import numpy as np
from hobotan import *

# Define quantum bits
q = symbols_list([3, 4], 'q{}_{}')

# Integer Encoding
x = 0
y = 0
z = 0
for i in range(4):
    x += (2**i) * q[0][i]
    y += (2**i) * q[1][i]
    z += (2**i) * q[2][i]

# Objective : Pythagoras cost
H1 = (x**2 + y**2 - z**2) **2

# Constraint: For the qubits composing x, none of the qubits are set to 0. The same applies to y and z
H2 = 0
H2 += np.prod((1 - q[0, :]))
H2 += np.prod((1 - q[1, :]))
H2 += np.prod((1 - q[2, :]))

# Combine terms
H = H1 + 10*H2

# Compile into HOBO tensor
hobo, offset = Compile(H).get_hobo()
print(f'offset\n{offset}')

# Select sampler
solver = sampler.MIKASAmpler()

# Perform sampling
result = solver.run(hobo, shots=10000)

# Display top 5 results
for r in result[:5]:
    print(f'Energy {r[1]}, Occurrence {r[2]}')
    
    # Display the result for each integer
    print("x =", Auto_array(r[0]).get_nbit_value(x))
    print("y =", Auto_array(r[0]).get_nbit_value(y))
    print("z =", Auto_array(r[0]).get_nbit_value(z))
\end{lstlisting}

\subsubsection{Results}

\begin{verbatim}
offset
30.0
MODE: GPU
DEVICE: cuda:0
Energy -30.0, Occurrence 480
x = 8.0
y = 6.0
z = 10.0
Energy -30.0, Occurrence 842
x = 4.0
y = 3.0
z = 5.0
Energy -30.0, Occurrence 1430
x = 12.0
y = 9.0
z = 15.0
Energy -30.0, Occurrence 1541
x = 12.0
y = 5.0
z = 13.0
Energy -30.0, Occurrence 150
x = 6.0
y = 8.0
z = 10.0
\end{verbatim}

The program above defines a function to create custom binary variables and constraints to ensure non-zero values. It then sets up the HOBO problem to find Pythagorean triples, compiles it into a HOBO tensor, and uses the MIKASAmpler to find solutions. The results show the first 8 solutions found, along with their corresponding energies and occurrences, and the execution time.

\subsection{Problem 3: Solving the Traveling Salesman Problem using Integer Encoding}

\subsubsection{Problem Formulation}

In this problem, we solve the Traveling Salesman Problem (TSP) using integer encoding to represent the path and distance~\cite{Glos2022}. The objective is to find the shortest possible route that visits each city once and returns to the origin city. While binary encoding typically uses one-hot constraints, integer encoding has been shown to significantly reduce the number of required qubits, allowing for more efficient solutions.

In the traditional TSP, the order of visiting cities is represented by one-hot vectors. However, with integer encoding, the sequence of visiting cities is represented as integers from 0 to 3. For this implementation, we set constraints such that the product of the integers representing the sequence of the 1st, 2nd, and 3rd cities equals 6, following the constraints from existing literature.

\subsubsection{Implementation}

\begin{lstlisting}[language=Python]
import numpy as np
from hobotan import *

# Define quantum bits
q = symbols_list([4, 2], 'q{}_{}')
xB = 2*q[0][0] + q[0][1]
xC = 2*q[1][0] + q[1][1]
xD = 2*q[2][0] + q[2][1]

H1 = 0  # This time, we will not set a cost

H2 = (xB*xC*xD-6)**2

# Combine terms
H = H1 + 10*H2

# Compile into HOBO tensor
hobo, offset = Compile(H).get_hobo()
print(f'offset\n{offset}')

# Select sampler
solver = sampler.MIKASAmpler()

# Perform sampling
result = solver.run(hobo, shots=10000)

# Display top 3 results
for r in result[:3]:
    print(f'Energy {r[1]}, Occurrence {r[2]}')

    # Display the result for each number
    print("xB =", Auto_array(r[0]).get_nbit_value(xB))
    print("xC =", Auto_array(r[0]).get_nbit_value(xC))
    print("xD =", Auto_array(r[0]).get_nbit_value(xD))
\end{lstlisting}

\begin{verbatim}
offset
360.0
MODE: GPU
DEVICE: cuda:0
Energy -360.0, Occurrence 865
xB = 1.0
xC = 2.0
xD = 3.0
Energy -360.0, Occurrence 1886
xB = 1.0
xC = 3.0
xD = 2.0
Energy -360.0, Occurrence 780
xB = 2.0
xC = 1.0
xD = 3.0
\end{verbatim}

We were able to solve the TSP efficiently by using integer encoding for the city visit sequences, significantly reducing the number of required qubits and improving computational efficiency.

\section{Optimizing HOBO Tensor}

In this section, we discuss the optimization of the HOBO tensor and the use of tensor networks for contraction and decomposition. The optimization of large tensors is crucial because they can be computationally intensive. By carefully designing the contraction path, we can achieve faster computations.

\subsection{Tensor Network Contraction}

Tensor networks are graphical representations of tensor contractions. Each node represents a tensor, and each edge represents a contraction between two tensors. The goal is to contract all tensors to obtain a final scalar or reduced tensor.

For example, consider the contraction of three tensors \(A\), \(B\), and \(C\):

\[
D = \sum_{i,j,k} A_{ij} B_{jk} C_{ki}
\]

In this case, the order of contractions significantly affects the computational complexity. We can optimize the contraction path to minimize the total number of operations.

\subsection{Optimizing Contraction Paths}

The choice of contraction path is crucial for efficient tensor computations. Different network shapes and configurations require different optimal paths. Tools like the \texttt{opt\_einsum} library in Python help find the best contraction path automatically.

For example, in the contraction of a high-dimensional tensor network, the optimal path can be chosen based on the following criteria:

\begin{itemize}
    \item \textbf{Minimize the intermediate tensor size}: Reduce the size of tensors generated during intermediate steps.
    \item \textbf{Reduce the total number of operations}: Minimize the computational cost by choosing the path with the least number of contractions.
\end{itemize}

By leveraging these optimization techniques, we can significantly improve the performance of the HOBO solver, making it capable of handling larger and more complex problems efficiently.

\subsection{Decomposing Large Tensors and Preparing for Path Optimization}

By decomposing large tensors, we can increase the options for selecting tensor contraction paths. Additionally, introducing approximations during decomposition can further reduce computational weight. In this section, we prepare for path optimization through decomposition without approximation. The primary decomposition method we will use is Singular Value Decomposition (SVD).

\subsection{Singular Value Decomposition (SVD)}

Singular Value Decomposition (SVD) is a method for decomposing any matrix \(A\) into the product of three simpler matrices:

\[
A = U \Sigma V^*
\]

Where:

\begin{itemize}
    \item \(A\) is an \(m \times n\) matrix.
    \item \(U\) is an \(m \times m\) unitary matrix.
    \item \(\Sigma\) is an \(m \times n\) diagonal matrix with singular values on the diagonal.
    \item \(V^*\) is the conjugate transpose of an \(n \times n\) unitary matrix.
\end{itemize}

This decomposition expresses the original matrix \(A\) as a product of three simpler matrices, simplifying tensor operations.

\subsection{Tensor Train Decomposition (TT Decomposition)}

Next, we use SVD consecutively to decompose a large tensor into a network of smaller tensors, known as Tensor Train Decomposition (TT Decomposition). TT Decomposition expresses a high-dimensional tensor as a sequence of 3-dimensional tensors.

When decomposing a high-dimensional tensor \(\mathcal{A}\) into a TT format, it can be represented as:

\[
\mathcal{A}_{i_1, i_2, \ldots, i_d} = \sum_{r_0, r_1, \ldots, r_{d-1}} 
G_1[r_0, i_1, r_1] \cdots G_d[r_{d-1}, i_d, r_d]
\]

Where:

\begin{itemize}
    \item \(\mathcal{A}\) is a \(d\)-dimensional tensor.
    \item \(G_k\) is a 3-dimensional tensor such that \(G_k \in \mathbb{R}^{r_{k-1} \times n_k \times r_k}\).
    \item \(r_0\) and \(r_d\) are virtual boundary conditions set to 1.
\end{itemize}

This TT Decomposition is performed sequentially as follows:

\begin{enumerate}
    \item Reshape the high-dimensional tensor \(\mathcal{A}\) into a matrix \(A_{(1)}\).
    \item Apply SVD to \(A_{(1)}\), decomposing it into \(U \Sigma V^*\).
    \item Incorporate \(\Sigma V^*\) into the next tensor and reshape it into a matrix \(A_{(2)}\).
    \item Repeat this process until the \(d\)-dimensional tensor is fully decomposed.
\end{enumerate}

Specifically, the process can be written as:

\[
A_{(1)} \approx U_1 \Sigma_1 V_1^*
\]
\[
A_{(2)} = \Sigma_1 V_1^*
\]
\[
A_{(2)} \approx U_2 \Sigma_2 V_2^*
\]

By repeating this process, a large tensor is decomposed into a network of smaller tensors. This approach increases the options for contraction paths and lays the groundwork for optimization.

\subsection{Solving TSP Using TT Decomposition}

In this subsection, we explore solving the Traveling Salesman Problem (TSP) using Tensor Train (TT) decomposition. By applying TT decomposition, we can significantly reduce the computational complexity of the problem.

\subsubsection{Before Decomposition}

In the TSP problem, we used the same formulation as in the previous section's \textit{Problem 3: Solving the Traveling Salesman Problem using Integer Encoding}. The size of the HOBO tensor was \((6,6,6,6,6,6)\).

This is the result of the numpy einsum path when contracting this tensor directly with the vector \( x \) to obtain the answer:

\begin{verbatim}
  Complete contraction:  i,j,k,l,m,n,ijklmn->
         Naive scaling:  6
     Optimized scaling:  6
      Naive FLOP count:  3.266e+05
  Optimized FLOP count:  1.120e+05
   Theoretical speedup:  2.917
  Largest intermediate:  7.776e+03 elements
\end{verbatim}

\begin{itemize}
    \item \textbf{Complete contraction:} All indices \(i, j, k, l, m, n\) are contracted to obtain the final scalar value.
    \item \textbf{Naive scaling:} The scaling of the calculation using a naive approach is 6. This indicates the computational complexity without any optimization.
    \item \textbf{Optimized scaling:} The scaling of the calculation using an optimized approach is also 6. This indicates the computational complexity after optimization, showing no reduction in this case.
    \item \textbf{Naive FLOP count:} The number of floating-point operations (FLOPs) in a naive approach is \(3.266 \times 10^5\). This measures the computational effort required without optimization.
    \item \textbf{Optimized FLOP count:} The number of FLOPs in an optimized approach is \(1.120 \times 10^5\). This shows the computational effort required after optimization, indicating a significant reduction.
    \item \textbf{Theoretical speedup:} The theoretical speedup achieved by optimization is approximately 2.917 times. This is the ratio of naive FLOP count to optimized FLOP count, indicating the efficiency gained.
    \item \textbf{Largest intermediate:} The largest intermediate tensor has \(7.776 \times 10^3\) elements. This represents the size of the largest tensor generated during the contraction process.
\end{itemize}

\subsubsection{After Decomposition}

This time, we performed TT decomposition on the TSP problem. Initially, the order-6 tensor of size \((6,6,6,6,6,6)\) was decomposed into six tensors with dimensions \([(6, 2), (2, 6, 3), (3, 6, 4), (4, 6, 4), (4, 6, 2), (2, 6)]\). Now, we will use these tensors to perform contraction with the vector \( x \).

This time, we performed TT decomposition on the TSP problem. Initially, the order-6 tensor of size \((6,6,6,6,6,6)\) was decomposed into six tensors with dimensions \([(6, 2), (2, 6, 3), (3, 6, 4), (4, 6, 4), (4, 6, 2), (2, 6)]\). Now, we will use these tensors to perform contraction with the vector \( x \).

The result of the numpy einsum path after performing the TT decomposition and contracting the resulting tensors with the vector \( x \) is as follows:

\begin{verbatim}
Complete contraction:  i,j,k,l,m,n,iA,AjB,
                       BkC,ClD,DmE,En->
       Naive scaling:  11
   Optimized scaling:  3
    Naive FLOP count:  1.075e+08
Optimized FLOP count:  7.010e+02
 Theoretical speedup:  153345.826
Largest intermediate:  1.600e+01 elements
\end{verbatim}

As a result, the maximum scaling parameter is limited to 3, significantly reduced from 6 before the decomposition. When performing TT decomposition, we did not apply any approximations and only optimized the computation path. As we can see, the computation has become significantly lighter after performing TT decomposition.

\section{Conclusion}

In this study, we introduced HOBOTAN, a Higher Order Binary Optimization (HOBO) solver that leverages tensor networks and PyTorch to efficiently solve combinatorial optimization problems. Throughout the paper, we demonstrated the capabilities of HOBOTAN by solving various complex problems, including the seating arrangement problem, the search for Pythagorean triples, and the Traveling Salesman Problem (TSP) using integer encoding.

By utilizing integer encoding, we significantly reduced the number of required qubits compared to traditional binary encoding methods, such as one-hot encoding. This reduction in qubits allowed for more efficient computation and optimized resource usage. Specifically, in the context of the TSP, the integer encoding of city visit sequences proved to be a powerful approach. This method enabled us to solve the TSP more efficiently by representing paths and distances in a compact form, thereby minimizing the computational overhead.

The incorporation of tensor networks into the HOBOTAN framework allowed us to efficiently manage and optimize high-dimensional tensors. By carefully designing contraction paths and employing decomposition techniques such as Singular Value Decomposition (SVD) and Tensor Train Decomposition, we were able to significantly enhance the computational efficiency of the solver. These optimizations are critical in handling large-scale problems that would otherwise be infeasible with conventional methods.

Furthermore, the integration of PyTorch provided powerful GPU acceleration capabilities, which are essential for handling the parallel computations involved in tensor contractions. The tensor batch system, introduced in HOBOTAN, further enhanced the solver's performance by enabling simultaneous computation of multiple vector contractions. This system proved particularly effective in real-world scenarios, allowing HOBOTAN to solve up to 10,000 samples rapidly and efficiently.

In conclusion, HOBOTAN demonstrates a significant advancement in solving higher-order binary optimization problems by leveraging the combined strengths of tensor networks and PyTorch. The solver's ability to efficiently handle complex problems with reduced computational resources makes it a promising tool for a wide range of applications in combinatorial optimization. Future work will focus on expanding HOBOTAN's capabilities, including multi-GPU support and further optimization of tensor network techniques, to tackle even larger and more complex problems.

\section{Discussion}

The current implementation of the HOBO solver has proven effective in solving complex optimization problems, confirming its utility. Given that it is based on tensor networks, a technology not yet fully exploited in this field, further exploration of this technology is necessary to enhance both speed and utility. Additionally, the continuous advancements in high-performance hardware offer a promising outlook for leveraging problem size and speed.

While the basic implementation of HOBOTAN includes numerous optimizations, there remains room for future improvements, such as supporting multi-GPU setups to further enhance computational capabilities. This opens up avenues for scaling the solver to tackle even larger and more complex problems efficiently.

\bibliography{apssamp}

\begin{thebibliography}{6}%
\makeatletter
\providecommand \@ifxundefined [1]{%
 \@ifx{#1\undefined}
}%
\providecommand \@ifnum [1]{%
 \ifnum #1\expandafter \@firstoftwo
 \else \expandafter \@secondoftwo
 \fi
}%
\providecommand \@ifx [1]{%
 \ifx #1\expandafter \@firstoftwo
 \else \expandafter \@secondoftwo
 \fi
}%
\providecommand \natexlab [1]{#1}%
\providecommand \enquote  [1]{``#1''}%
\providecommand \bibnamefont  [1]{#1}%
\providecommand \bibfnamefont [1]{#1}%
\providecommand \citenamefont [1]{#1}%
\providecommand \href@noop [0]{\@secondoftwo}%
\providecommand \href [0]{\begingroup \@sanitize@url \@href}%
\providecommand \@href[1]{\@@startlink{#1}\@@href}%
\providecommand \@@href[1]{\endgroup#1\@@endlink}%
\providecommand \@sanitize@url [0]{\catcode `\\12\catcode `\$12\catcode `\&12\catcode `\#12\catcode `\^12\catcode `\_12\catcode `\%12\relax}%
\providecommand \@@startlink[1]{}%
\providecommand \@@endlink[0]{}%
\providecommand \url  [0]{\begingroup\@sanitize@url \@url }%
\providecommand \@url [1]{\endgroup\@href {#1}{\urlprefix }}%
\providecommand \urlprefix  [0]{URL }%
\providecommand \Eprint [0]{\href }%
\providecommand \doibase [0]{https://doi.org/}%
\providecommand \selectlanguage [0]{\@gobble}%
\providecommand \bibinfo  [0]{\@secondoftwo}%
\providecommand \bibfield  [0]{\@secondoftwo}%
\providecommand \translation [1]{[#1]}%
\providecommand \BibitemOpen [0]{}%
\providecommand \bibitemStop [0]{}%
\providecommand \bibitemNoStop [0]{.\EOS\space}%
\providecommand \EOS [0]{\spacefactor3000\relax}%
\providecommand \BibitemShut  [1]{\csname bibitem#1\endcsname}%
\let\auto@bib@innerbib\@empty
\bibitem [{\citenamefont {Domino}\ \emph {et~al.}(2022)\citenamefont {Domino}, \citenamefont {Kundu}, \citenamefont {Salehi},\ and\ \citenamefont {Krawiec}}]{Domino_2022}%
  \BibitemOpen
  \bibfield  {author} {\bibinfo {author} {\bibfnamefont {K.}~\bibnamefont {Domino}}, \bibinfo {author} {\bibfnamefont {A.}~\bibnamefont {Kundu}}, \bibinfo {author} {\bibfnamefont {Ã.}~\bibnamefont {Salehi}},\ and\ \bibinfo {author} {\bibfnamefont {K.}~\bibnamefont {Krawiec}},\ }\bibfield  {title} {\bibinfo {title} {Quadratic and higher-order unconstrained binary optimization of railway rescheduling for quantum computing},\ }\bibfield  {journal} {\bibinfo  {journal} {Quantum Information Processing}\ }\textbf {\bibinfo {volume} {21}},\ \href {https://doi.org/10.1007/s11128-022-03670-y} {10.1007/s11128-022-03670-y} (\bibinfo {year} {2022})\BibitemShut {NoStop}%
\bibitem [{\citenamefont {Suksmono}\ and\ \citenamefont {Minato}(2022)}]{Suksmono2022}%
  \BibitemOpen
  \bibfield  {author} {\bibinfo {author} {\bibfnamefont {A.~B.}\ \bibnamefont {Suksmono}}\ and\ \bibinfo {author} {\bibfnamefont {Y.}~\bibnamefont {Minato}},\ }\bibfield  {title} {\bibinfo {title} {Quantum computing formulation of some classical hadamard matrix searching methods and its implementation on a quantum computer},\ }\href {https://doi.org/10.1038/s41598-021-03586-0} {\bibfield  {journal} {\bibinfo  {journal} {Scientific Reports}\ }\textbf {\bibinfo {volume} {12}},\ \bibinfo {pages} {197} (\bibinfo {year} {2022})}\BibitemShut {NoStop}%
\bibitem [{\citenamefont {Minato}(2024)}]{minato2024tensornetworkbasedhobo}%
  \BibitemOpen
  \bibfield  {author} {\bibinfo {author} {\bibfnamefont {Y.}~\bibnamefont {Minato}},\ }\href {https://arxiv.org/abs/2407.16106} {\bibinfo {title} {Tensor network based hobo solver}} (\bibinfo {year} {2024}),\ \Eprint {https://arxiv.org/abs/2407.16106} {arXiv:2407.16106 [quant-ph]} \BibitemShut {NoStop}%
\bibitem [{\citenamefont {Wang}\ \emph {et~al.}(2023)\citenamefont {Wang}, \citenamefont {Pan}, \citenamefont {Xu}, \citenamefont {Yang}, \citenamefont {Li},\ and\ \citenamefont {Cichocki}}]{wang2023tensornetworksmeetneural}%
  \BibitemOpen
  \bibfield  {author} {\bibinfo {author} {\bibfnamefont {M.}~\bibnamefont {Wang}}, \bibinfo {author} {\bibfnamefont {Y.}~\bibnamefont {Pan}}, \bibinfo {author} {\bibfnamefont {Z.}~\bibnamefont {Xu}}, \bibinfo {author} {\bibfnamefont {X.}~\bibnamefont {Yang}}, \bibinfo {author} {\bibfnamefont {G.}~\bibnamefont {Li}},\ and\ \bibinfo {author} {\bibfnamefont {A.}~\bibnamefont {Cichocki}},\ }\href {https://arxiv.org/abs/2302.09019} {\bibinfo {title} {Tensor networks meet neural networks: A survey and future perspectives}} (\bibinfo {year} {2023}),\ \Eprint {https://arxiv.org/abs/2302.09019} {arXiv:2302.09019 [cs.LG]} \BibitemShut {NoStop}%
\bibitem [{\citenamefont {Lucas}(2014)}]{Andrew2014}%
  \BibitemOpen
  \bibfield  {author} {\bibinfo {author} {\bibfnamefont {A.}~\bibnamefont {Lucas}},\ }\bibfield  {title} {\bibinfo {title} {Ising formulations of many np problems},\ }\href {https://arxiv.org/abs/1302.5843} {\bibfield  {journal} {\bibinfo  {journal} {arXiv preprint arXiv:1302.5843}\ } (\bibinfo {year} {2014})}\BibitemShut {NoStop}%
\bibitem [{\citenamefont {Glos}\ \emph {et~al.}(2022)\citenamefont {Glos}, \citenamefont {Krawiec},\ and\ \citenamefont {Zimborás}}]{Glos2022}%
  \BibitemOpen
  \bibfield  {author} {\bibinfo {author} {\bibfnamefont {A.}~\bibnamefont {Glos}}, \bibinfo {author} {\bibfnamefont {A.}~\bibnamefont {Krawiec}},\ and\ \bibinfo {author} {\bibfnamefont {Z.}~\bibnamefont {Zimborás}},\ }\bibfield  {title} {\bibinfo {title} {Space-efficient binary optimization for variational quantum computing},\ }\href {https://doi.org/10.1038/s41534-022-00546-y} {\bibfield  {journal} {\bibinfo  {journal} {npj Quantum Information}\ }\textbf {\bibinfo {volume} {8}},\ \bibinfo {pages} {39} (\bibinfo {year} {2022})}\BibitemShut {NoStop}%
\end{thebibliography}%

\end{document}